\documentclass[onecolumn,aps,preprint,showpacs,amsmath,amssymb]{revtex4}
\usepackage{graphicx}
\usepackage{latexsym}
\usepackage{xcolor} 
\usepackage{dcolumn} 
\begin{document}
\newcommand{\eq}{\begin{equation}}                                                                         
\newcommand{\eqe}{\end{equation}}             
 
\title{General self-similar solutions of diffusion equation and related constructions }

\author{Imre F. Barna$^{1}$ and L. M\'aty\'as$^{2}$}
\address{ $^1$ Wigner Research Center for Physics, 
\\ Konkoly-Thege Mikl\'os \'ut 29 - 33, 1121 Budapest, Hungary \\
$^2$  Department of Bioengineering, Faculty of Economics, Socio-Human Sciences
and Engineering, Sapientia Hungarian University of Transylvania
Libert\u{a}tii sq. 1, 530104 Miercurea Ciuc, Romania} 
\date{\today}

\date{\today}
\begin{abstract} 
Transport phenomena plays an important role in science and technology. In the wide variety of applications both advection 
and diffusion may appear. Regarding diffusion, for long times, different type of decay rates are possible for different 
non-equilibrium systems. After summarizing the existing solutions of the regular diffusion equation, we present not so well known 
solution derived from three different trial functions, as a key point we present a family of solutions for the case of infinite horizon. 
By this we tried to make a step toward understanding the different long time decays for different diffusive systems.      
\end{abstract}
\maketitle
\section{introduction}

The process of spreading is a quite common and relatively often occurring phenomena. 
If the process is sufficiently slow and exhibits certain features then it is called as a diffusion process. 
When the particle diffuses into it's own environment than it is called self-diffusion. 
This kind of diffusion is characterized often by different parameters in different regions of space. 

 This fundamental question became an enormous scientific field in the last century a historical review can be found 
in \cite{hist1,hist2}.  Beyond the most studied regular case there are non-linear (or anomalous)  
\cite{non-linear, degen,anom},  non-local  \cite{non-loc} or even fractional processes  \cite{frac-dif}. 
  Far from completeness we just mention some of the most relevant monographs. In the following we deal with the 
regular diffusion equation therefore we mention three basic references \cite{crank, Balescu1997, sush}.  
It is obvious that this phenomena has an important number of applications in  engineering 
\cite{Bennett2013,Bird2015}, in meteorology \cite{pasquil,envir} in  polymer science \cite{poly}, in finance \cite{fin} 
or even in social networks \cite{soc}. 

Beyond the phenomenological macroscopic description of diffusion there are numerous 
models exist to study the process. 
For instance the inhomogeneity in a gas means a non-constant density, and usually also an inhomogeneous pressure. 
General equations of motion of the one component fluid can be found in \cite{Balescu1997}.
The case of binary diffusion means a diffusion of given particles among other ones which are different. 
The binary diffusion has interesting properties and eventually spectacular in case if it is visible. 
Regarding models of binary diffusion, it is worthwhile to mention the Lorentz gas model. 
In this model certain results have been obtained related to diffusion by Machta and Zwanzig \cite{MaZw}, 
Claus and Gaspard \cite{ClGa2001}. 
Connections with non-equilibrium thermodynamics were analyzed in \cite{GaNiDo2002}, and \cite{MaTeVo2004}. 
Beyond diffusion M\'aty\'as and Gaspard \cite{MaGa2005} discussed diffusion with a simple reaction - isomerization -  
where not only the diffusion coefficient, but the reaction rate is also evaluated. 
One may also find diffusive processes where the diffusion is determined by the surroundings or the boundaries or the shape of the surface \cite{MaKl2004,KlBaMa2004,MaBa2011,HaHa2018}.

Regarding biological applications, the diffusion equation has an important role at mesoscopic - cell size – 
scale \cite{AnKo2015}, and also in the design of bioreactors \cite{Panda}. 
 
 Diffusive aspects one may find in certain hydrodynamic equations with dissipation \cite{If2002,Hu2010,Ji2009,Fa2010,Ba2011,YeYu2011,BaMa2014} or in quantum related systems \cite{Wu2010,Dai2010,Gao2009}. 

We find it important to emphasize that the free-particle Schr\"odinger equation -- from mathematical viewpoint -- is also a kind of diffusion equation
 \cite{schro1,schro2} therefore all the forthcoming analysis could lead to reasonable results for that equation as well.   

It is obvious, but we have to state that mathematically heat conduction is 
similar to diffusion. The field has its mighty literature as well from which we mention two recent monographs \cite{heat1,heat2}.   
Regarding numerical methods is worthwhile to mention the solutions obtained in \cite{Ko2020a,Ko2020b}. 

Similar equations which may take into account certain other perturbations in the system are the telegraph 
equation which is "obviously hyperbolic" \cite{jos} or other telegraph-type equations 
like the Euler-Poisson-Darboux which can be derived from the modified Fick's (or Fourier's) law \cite{barna1,garra}. 
 Another answer is the investigation of the nonlinear hyperbolic system of diffusion flux relaxation and 
energy conservation equations instead of the second order diffusion equation \cite{imre-robi1,imre-robi2}.  Such first order system may have 
shock-wave characters as well.  
The literature of this question is again numerous and we do not go into further details. 

As final point we should mention scientific research fields and mathematical problems which grew out of the original diffusion 
problem such as reaction-diffusion \cite {reacdif1,reacdif2}, porous media studies \cite{porous},  surface growth 
phenomena \cite{kpz},  fractional 
diffusion \cite{fractional} or p-Laplacian \cite{plaplace1,plaplace2} equations.  

\section{Theory and Results}
Having in mind that the general diffusion process is three dimensional we consider only 
 one Cartesian coordinate, therefore the equation reads  
\begin{eqnarray}
\frac{\partial C(x,t)}{\partial t  } = D \frac{\partial^2 C(x,t)}{\partial x^2  }, 
\label{pde} 
\end{eqnarray}
where  $C(x,t)$ is the distributions of the particle concentration in space and time and $D$  is the diffusion 
coefficient. $C(x,t)$ in the equation above is considered up to a constant, consequently it may also refer to the 
concentration above or around the average. The 
function $C(x,t)$ fulfills the necessary smoothness conditions with existing continuous first and second derivatives 
in respect to time and and space and from physical reasons $D > 0$. 
Numerous physics textbooks gives us the derivation how the fundamental (the Gaussian) solutions can be obtained e.g. \cite{crank,carl,math,tham}.  
 First, to dispel misunderstandings we have to express one thing clearly, the  
regular diffusion equation has existence and unicity theorem for initial and boundary problems, but this is not contradictory to our forthcoming analysis. We will apply three 
different trial functions (or Ans\"atze [this is the plural form] ) but neither the initial nor the boundary problems 
are being well defined. The obtained results may fulfill well-defined initial and boundary problems via fixing their integration constants $c_1$ and $c_2$.

 In 1969 Bluman and Cole  \cite{bluman} gave an analysis based on a general symmetry analysis method giving 
numerous analytic solutions, some of them are expressible with Gaussian or error functions.  At this generality, 
presented below, there was a need of almost all confluent hypergeometric functions, to describe the phenomena. 
In the following first we give some additional exact solutions of the diffusion equation ending up with an in-depth analysis of the classical self-similar 
solutions which can have physical applications.  

As note zero we must say that with trivial derivation all reader can verify that the functions
\eq
 \left\{t + \frac{Dx^2}{2}, \hspace*{3mm}  exp(t \mp \sqrt{D} x), \hspace*{3mm}  exp(-t)\cdot (cos[ \sqrt{D}  x] + sin[\sqrt{D} x]) \right\},  
\label{trivial}
\eqe 
 are all solutions of Eq (\ref{pde}). These are called separable solutions, the first one is an additive separable solution $C(x,t) = f(t) +g(x)$ and 
the following two are multiplicative separable  $C(x,t) = h(x) \cdot (t)$ solutions in respect to the spatial and temporal variables.)  
These solutions are usually mentioned in textbook analysis, and  will be relevant later on, as we will see.  
It is also interesting to note, that the traveling wave Ansatz -- which mimics the wave properties of the investigated PDE -- 
 $C(x,t) = f(x \mp ct)$ automatically gives the exponential solutions. 

Beyond the analysis of Bluman and Cole there is an other celebrated work of Clarkson and Kruskal \cite{krusk} describing the 
non-classical method of group invariant solutions, which is often used to obtain 
similarity solutions of (mostly non-linear) PDEs. Originally it was introduced and applied for the Boussinesq equation. 
Now we apply it to the diffusion equation. (Due to our best knowledge it was not done and not presented in a clear-cut way till now.)
The Ansatz has the form of 
\eq
C(x,t) =   \beta(x,t)\cdot W(z[x,t])  
\eqe
where all real functions $\alpha,\beta,W$ and $z$ should have existing first continuous derivatives in respect to time and the second 
existing continuous derivatives in respect to the coordinate  $x$, finally $W(z[x,t])$ is a compound function. 
The method to derive the solution is the following, the first temporal and second spatial derivatives have to be evaluated and replaced into 
Eq (\ref{pde}) giving 
\eq
  \beta_t W + \beta W' z_t = D(\beta_{xx}W + 2\beta W' z_x  +\beta_x W' z_x + \beta W'' (z_x)^2 + \beta W' z_{xx}),
\eqe
where the subscripts x and t mean partial derivation in respect to time and coordinate, and prime means derivation of $W(z)$ in respect to $z$. 
The key idea is the following we should like to have an ordinary differential equation (ODE) for $W$ as independent variable $z$ 
With reorganization of the terms it reads
\eq
  W'' (D \beta (z_x)^2) + W' (\beta z_t - 2 D \beta z_x - D\beta_x z_x - D \beta z_{xx}) + W(\beta_t - D\beta_{xx}) = 0. 
\label{odeW}
\eqe
To solve this expression as an ODE for $W(z)$ we have to fix that the factors of the second, first and zeroth derivative are real constants  
\begin{eqnarray}
 D \beta (z_x)^2 = \tilde{C}_1,  \label{eqn:1}  \\ 
\beta z_t - 2 D \beta z_x - D\beta_x z_x - D \beta z_{xx} = \tilde{C}_2,     \label{eqn:2} \\  
\beta_t - D\beta_{xx} = \tilde{C}_3, \hspace*{3mm}  W \ne 0.  \label{eqn:3}
\end{eqnarray} 
There are numerous ways to solve this system. 
Note, that  (\ref{eqn:3}) is identical to the original diffusion equation if $\tilde{C}_3 = 0$. Therefore if we know 
and kind of solution, (as starting point we may take the trivial solutions of 
Eq. (\ref{trivial}) -- which are additive or multiplicative solutions --) then simply integrating  Eqs. (\ref{eqn:1} - \ref{eqn:2}) 
and finally the ODE Eq. (\ref{odeW}) numerous solutions can be derived.  (Finally, it is important to note, that in the original 
paper of Clarkson and Kruskal there is an additional functions in the Ansatz 
$ C(x,t) =  \alpha(x,t) +  \beta(x,t)\cdot W(z[x,t])  $  
which is $\alpha(x,t)$ and  is important for non-linear PDE. However for the linear diffusion equation the superposition is valid and we can neglect it.)
These mathematically correct solutions become very compound and complicated 
if we start with the Gaussian solution and has  little physical interest. 

Before we get to our essential point we show a second kind of solution which -- in theory -- interpolates the traveling wave Ansatz $C(x,t) = g(x \mp c \cdot t) $ and  
the disperse self-similar Ansatz of $C(x,t) = t^{-\alpha}f(x/t^{\beta})$. This trial function is called traveling-profile 
function and was introduced by Benhamidouche \cite{behn} with the form of 
\eq
C(x,t) = a(t) \cdot h\left( \frac{x-b(t)}{c(t)} \right) =  a(t) \cdot h(\omega) , 
\eqe
where all $a,b,c$ and $h$ are continuous real functions with existing continuous first temporal and second spatial derivatives and $\omega$ is the new 
reduced independent variable.   
The solution method is similar as explained above.  Performing the spatial and temporal derivations and substitution back to (\ref{pde})  
we arrive to 
\eq
\frac{c^2 a_t}{a} h + a h' [b_t c - \eta c c_t] = D h'', \hspace*{1cm}  a(t)/c(t)^2 \ne 0,    
\label{f_trav_prof}
\eqe
where prime means derivation in respect to $\omega$ and subscript t in respect to time.  
This equation should be an ODE for $h(\omega)$ therefore the coefficients of $h$ and $h'$ should be independent of time, should be constants, therefore 
the following constraints have to be fulfilled: 
\eq
\frac{c^2 a_t}{a} = \tilde{C}_1, \hspace*{1cm}
b_t  c          = \tilde{C}_2, \hspace*{1cm}
 c_t c                 = \tilde{C}_3.  
\eqe
All three solutions can be easily obtained by direct integration, (starting with the last equation) and read 
\eq
c(t) = \mp \sqrt{2 \tilde{C}_3 t +c_1},  \hspace*{2mm} b(t) = \frac{\mp \sqrt{2 \tilde{C}_3 t +c_1}  }{ \tilde{C}_3} + c_2,
 \hspace*{2mm} a(t) = c_3 (2 \tilde{C}_3 t +c_1)^{\frac{\tilde{C}_1}{2 \tilde{C}_3 }},  
\eqe
finally the solution of the traveling profile shape function is 
\eq
h(\omega) = c_4M\left(- \frac{\tilde{C}_1}{2\tilde{C}_3},\frac{1}{2}, \frac{[\tilde{C}_2 -  \tilde{C}_2\omega ]^2 }{D \tilde{C}_3}   \right)   
+ c_5 U \left(- \frac{\tilde{C}_1}{2\tilde{C}_3},\frac{1}{2}, \frac{[\tilde{C}_2 -  \tilde{C}_2\omega ]^2 }{D \tilde{C}_3}   \right),   
\label{travp}
\eqe
where M and U are  the Kummer functions \cite{NIST} 
with the argument of 
\eq
\omega = \frac{x-b(t)}{c(t)} = \frac{x}{ \sqrt{2 \tilde{C}_3 t +c_1} }  - \frac{c_2}{  \sqrt{2 \tilde{C}_3 t +c_1} }- \frac{}{} \frac{ \tilde{C}_2 }{\tilde{C}_3}.
\eqe

Finally, we concentrate on our main Ansatz, 
on the self-similar one 
\begin{eqnarray}
C(x,t) = t^{-\alpha} f\left(\frac{x}{t^{\beta}} \right)  = t^{-\alpha} f(\eta)  \hspace*{3mm}  
\label{ansatz}
\end{eqnarray}
where $\alpha$ and $\beta$ are the self-similar exponents being real numbers describing the decay and the spreading of the solution is time and space.  
These properties makes this Ansatz physically extraordinary relevant and was first introduced by 
Sedov \cite{sedov} later used by Zel'dowich and Raizer \cite{zeld} and Barenblatt \cite{barenb}. 
In the last decade we applied this trial function to numerous non-linear PDE systems, most of them are from fluid dynamics \cite{imre1} but 
investigated electromagnetic \cite{imre2}  or quantum mechanical problems \cite{imre3} as well. 
 The self-similar analysis have been successfully applied to different systems, where diffusion may also appear \cite{NaSi19,Sa20,KaSuZh20}.    

The forthcoming analysis is quite simple, and similar to the former ones. Let's calculate the first time 
and second spatial derivative of the Ansatz (\ref{ansatz}) using the derivation rule of the indirect function and put in into the diffusion equation 
(\ref{pde}), we arrive at 
\eq
-\alpha t^{-\alpha -1} f(\eta)  - \beta   t^{-\alpha -1} \eta  
\frac{d f(\eta)}{d\eta}  =  D  t^{-\alpha -2\beta}  \frac{d^2f(\eta)}{ d\eta^2}.   
\eqe
Now comes the crucial point of the reduction mechanism (or the applicability of the Ansatz) if all three 
terms has the same time dependence, (all exponents are the same) then all can be canceled by  an 
algebraic simplification and a clear-cut ODE is derived for the shape function.  So the relation among the all time dependent factors (now only two)  
\eq
t^{-\alpha -1} \hspace*{0.5cm}  \stackrel{?}{=}  \hspace*{0.5cm}  t^{-\alpha -2\beta},
\label{how}
\eqe
have to investigated. At this point we have to mention that this analysis 
can be generalized for PDE systems with 4-5 variables even for multiple spatial dimensions  as 
well, e.g. \cite{imre1,imre2,imre3,imre4}  which makes the method very striking and effective. Compared to the general 
Lie symmetry analysis the method remains transparent even for a PDE system of 4-5 variables. 
The analysis of the relations among the self-similar exponents (now for $\alpha,\beta$) can end up with three different scenarios:
\begin{itemize}
\item{
The linear algebraic equation system among the exponents can be overdetermined, 
which automatically means contradiction. Therefore the system has inherently no physically 
self-similar power-law decaying or exploding solutions. Such systems are rare but some damped wave equations e.g. telegraph equations are so.} 
\item{ All exponents have well-defined numerical values, the analysis of the solutions is 
straightforward, the remaining coupled non-linear ODE system can be analyzed, in some lucky 
cases even it can be decoupled and in best cases all variables can be expressed with analytic formulas. Such a system is the incompressible Navier-Stokes 
equation \cite{Ba2011}  where all exponents have the same numerical value of 1/2, except the time decay of the pressure function which is +1.}  
\item{The linear algebraic equation system for the exponents are under-determined, 
leaving usually one self-similar exponent completely free, which means an extra free parameter in the 
obtained ODE system, causing a very rich mathematical structure. The free exponent can have either positive or 
negative sign. Negative values usually result in power-law divergent or exploding solutions in contrary, positive 
exponents mean power-law decaying solutions which are desirable for dissipative systems. 
 This is the case for the present regular diffusion equation. }
\end{itemize}

For the present diffusion equation, assuming that in (\ref{how}) the equality strictly holds, after some trivial algebra we get: 
\eq 
\alpha =     \textrm{arbitrary real number},   \hspace*{3mm} \beta = 1/2,
\eqe
there is a clear-cut time-independent ODE of 
\eq
-\alpha f - \frac{1}{2}\eta f' = D f''.  
\label{ode1}
\eqe
With $ \alpha = 1/2 $ the left-hand side of ODE is  total derivative and can be integrated 
getting 
\eq
- \frac{1}{2}\eta f +c_1 = D f',  
\eqe
if the integration constant -- which can be interpreted as a source term-- is taken to be zero $c_1 =0 $ 
we get back the usual Gaussian solutions of 
\eq
f(\eta) = c_2 e^{-\frac{\eta^2}{4D}},
\eqe 
from the final solution of 
\eq
C(x,t) = c_2 t^{-1/2}  e^{-\frac{x^2}{4Dt}},
\label{Gauss}
\eqe 
we can read, that $\beta$ is responsible for the spreading 
and $\alpha$ is for the decay of the solution. 
This is a general, (and very powerful) feature of the self-similar Ansatz 
that positive $\alpha$ and $\beta$ values always represent decaying and 
spreading solutions, which have great physical relevance.  
(This fundamental solution is sometimes referred to as $ {\it{source \> type}}$ solution -- by 
mathematicians --  because for $t \rightarrow 0 $  then $C(x,0) \rightarrow  \delta(x)$.) 

We will see later on that,  $\alpha < 0$ values mean exploding solutions which have only mathematical interest in most cases. 
 It is also clear from (\ref{Gauss}) that no real solutions can be defined for $t<0$.The spreading and decaying properties are visualized on Fig. 1 below.  
\begin{figure}[!h]
\scalebox{0.5}{
\rotatebox{0}{\includegraphics{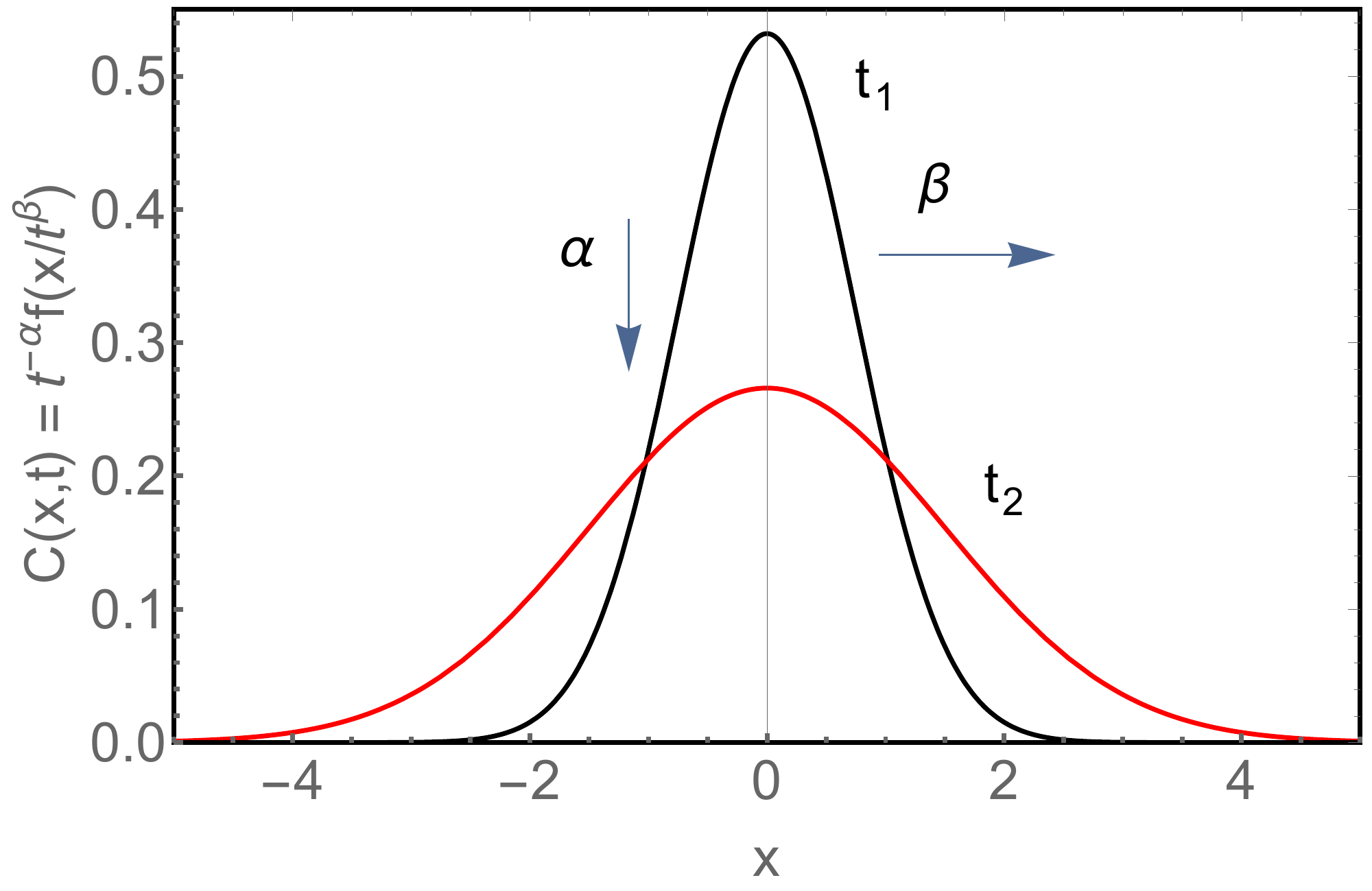}}}
\caption{A self-similar solution of Eq. (\ref{pde}) for $t_1<t_2$.
The presented curves are the Gaussians. The physical role of the self-similar exponents 
are indicated with arrows.}	
\label{egyes}       
\end{figure}
 
\begin{figure}  
\scalebox{0.4}{
\rotatebox{0}{\includegraphics{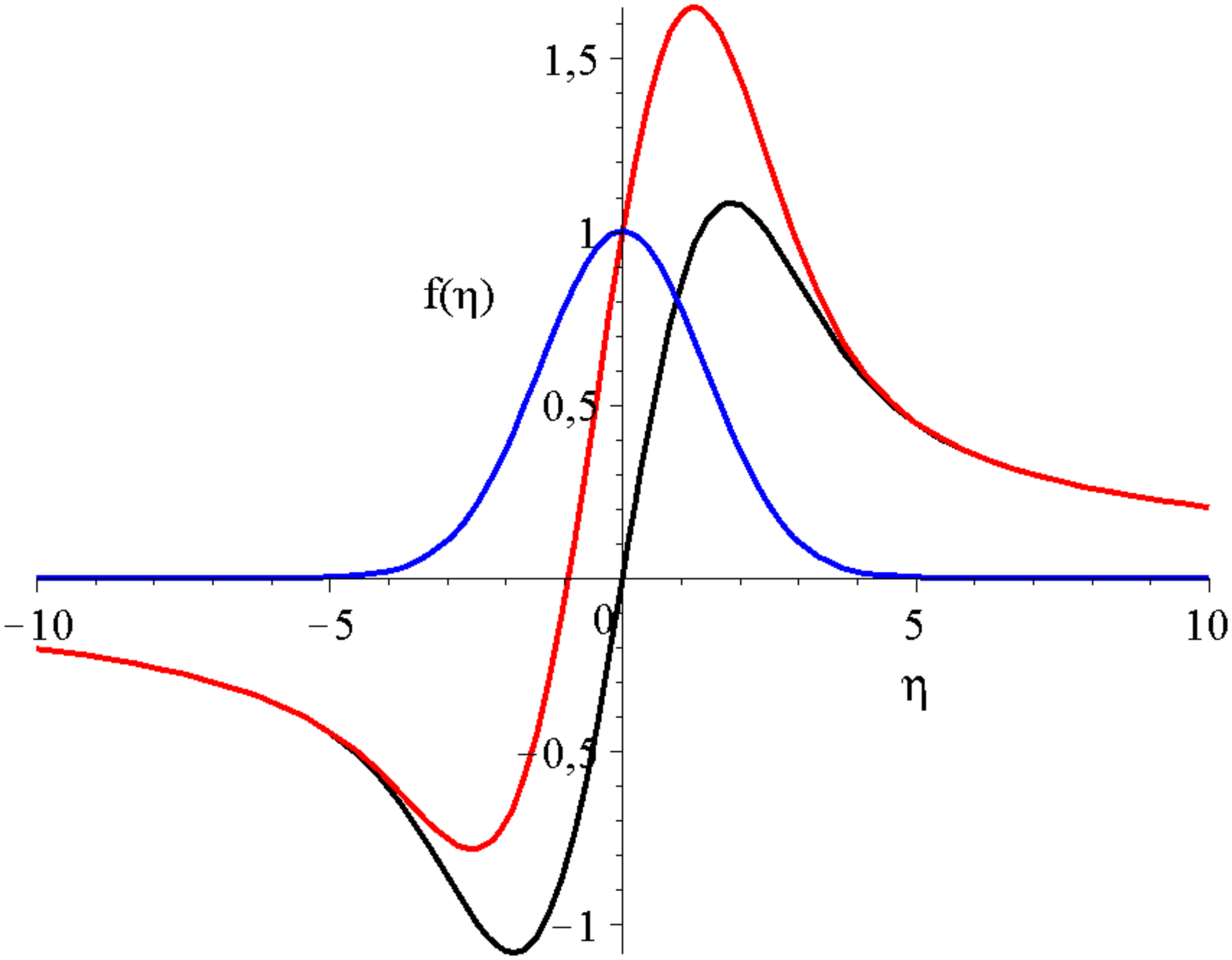}}}
\caption{Numerous shape functions  $f(\eta)$ Eq. (\ref{f0}) for three different initial conditions  the black,red and 
blue curves are for $c_1 = 1, c_2 =0$, $c_1 = c_2 = 1$ and for $c_1 = 0, c_2 = 1$, respectively.}
\label{kettes}      
\end{figure}

If the more general $c_1 \ne 0$ is taken 
then the solutions is changed to 
\eq
f(\eta) = \left(  \frac{c_1 \sqrt{\pi} \cdot erf \left[ \frac{1}{2} \sqrt{- \frac{1}{D}}  \eta  \right]    }
{D\sqrt{-\frac{1}{D}} }    + 
  c_2  \right) \cdot e^{-\frac{\eta^2}{4D}},
\label{f0}
\eqe
 where erf is the error function \cite{NIST}.  (We present the formal solutions obtained by the Maple 12 Software 
 [Copyright (c) Maplesoft, a division of Waterloo Inc. 1981 -2008] from now on.)
This solutions is not so commonly known. Non zero $c_1$ integration constant modifies the shape of the Gaussian solution. 
For clarity, figure (\ref{kettes}) shows the solutions for different initial conditions. 
 
The second -- and more general -- case is for $\alpha \ne 1/2$, 
($\beta$ is still  one half) now the solution of Eq.  (\ref{ode1}) reads:
\eq
f(\eta) = \eta  \cdot e^{-\frac{ \eta^2}{4D} }  \left(  c_1 M\left[1-\alpha  , \frac{3}{2} , \frac{ \eta^2}{4D} \right]  + c_2 U\left[1 -\alpha , \frac{3}{2} , \frac{ \eta^2}{4D} \right]    \right),
\label{f_eta2}
\eqe
 where  $M(\cdot,\cdot,\cdot) $ and $U(\cdot,\cdot,\cdot) $ are the Kummer's functions for exhaustive details see the NIST Handbook \cite{NIST}.   
\begin{figure}  
\scalebox{0.4}{
\rotatebox{0}{\includegraphics{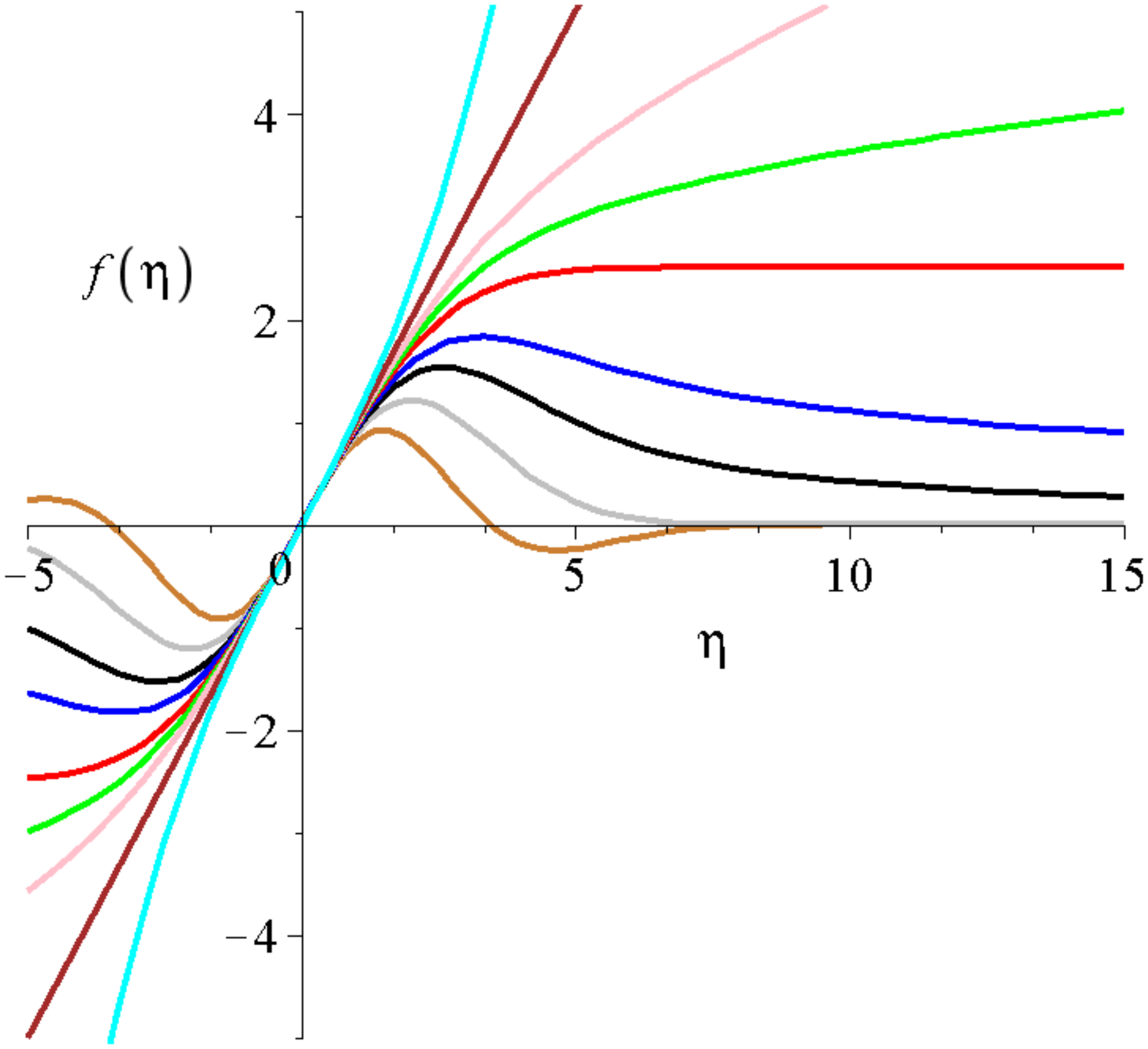}}}
\caption{Numerous shape functions  $f(\eta)$ Eq. (\ref{f_eta2}) for various $\alpha$s all are for $\beta = 1/2$ and for $c_1 = 1, c_2 =0, D =2	$.
The gold, gray, black, blue, red, green, pink, brown and cyan curves are for $\alpha = 2,1,1/2,1/4,0,-1/8,-1/4,-1/2$ and for $-1$, respectively.}
\label{harmas}      
\end{figure}

For $\alpha > 0$ the solution tends to zero for large values of $\eta$. 

For $\alpha< 0$ one can see divergent  solutions, which go to infinity at 
infinite argument. (The blow-up type of solutions are different and will be defined later.)
 From the series expansion of M 
we get 
\begin{equation}
M(a,b,z) = 1 + \frac{az}{b} + \frac{(a)_2 z^2}{(b)_2 2!} + ... + 
  \frac{(a)_n z^n}{(b)_n n!}, 
\end{equation}
with the $(a)_n = a(a+1)(a+2)...(a+n-1), (a) _0 = 1 $ is the so-called 
rising factorial or Pochhammer's Symbol \cite{NIST}. 
In our present case $b$ has a fix non-negative integer value, so none of the solutions have poles at $b = -n$. 
For the Kummer function $M$ when the parameter $a$ has negative integer numerical values  ($a = -m$)  the solution is reduced to 
 a polynomial of degree $m$ for the variable $z$.  In other cases $a \ne -m$ we get a convergent infinite series for all values of $a,b$ and $z$.  
There is a connecting formula between the two Kummer functions,   
$U$ is defined from $M$ via 
\eq
U(a,b,z) = \frac{\pi}{sin(\pi b)} 
\left( \frac{M[a,b,z]}{\Gamma[1+a-b]\Gamma[b]} 
- z^{1-b}\frac{M[1+a-b,2-b,z]}{\Gamma[a]\Gamma[2-b]} \right),     
\eqe
where $\Gamma(a)$ is the Gamma function \cite{NIST}.
Figure 3 shows the shape functions for numerous different values of $\alpha$. Note, that all positive $\alpha$s mean solution with asymptotic 
decay which means that for $\eta \rightarrow \infty $ the $f(\eta) \rightarrow 0$. 
This can interpreted as certain kind of boundary conditions $f(0) = 0$ and $f(\eta \rightarrow \infty) \rightarrow 0 $.   
The $ \alpha > 0$ means additional oscillations. 
Zero alpha value means a solution which converges to a finite value, 
and negative values  means divergent solutions. 

 Note, the clear difference between the two solutions obtained from the 
traveling profile (\ref{travp})  and the self-similar Ansatz (\ref{f_eta2}). Both contain 
Kummer functions but with different arguments and coefficient functions. 
  
As one can see, $\alpha = 0$ is a special case, when Eq. (\ref{ode1}) is simplified to 
\eq
-\frac{\eta f'}{2} = Df'',  
\eqe
resulting 
\eq
f(\eta) = c_1 + c_2 \cdot erf\left( \sqrt{\frac{\beta}{ 2D}} \eta  \right),
\eqe
which is a sigmoid function which starts from zero and tends to a nonzero constant at large values argument $\eta$.  
As a consequence $C(x,t)$ also tends to a constant for these values. 

From practical point of view, this case has certain similarities with the evaporation phenomena  \cite{evap}, when initially 
there is no vapor concentration above the liquid, and as the time passes above the liquid phase vapor appears, which becomes denser with time. 

  If $\alpha=1$ we have an interesting case. 
The first argument of the Kummer functions $M$ and $U$ is $1-\alpha = 0$. 
This means, that 
\begin{equation} 
M \left(0,\frac{3}{2},\frac{\eta^2}{4D}\right) = 1,   
\end{equation}   
and the other function $U(0,3/2,\eta^2/[4D])$ is also constant. 
Consequently the general solution is  
\begin{equation} 
C(x,t) = \frac{1}{t} f(\eta) =  \frac{1}{t} \eta e^{-\frac{\eta^2}{4D}} \cdot Const. 
\end{equation}   

Case $\alpha=2$ yields more in the expression of the function $f(\eta)$. 
The first argument of the functions $M$ and $U$ is $1-\alpha = -1$. 
In this case the function $M$ is a first order polynomial, the higher order coefficients vanishes,  
\begin{equation} 
M\left(-1,\frac{3}{2},\frac{\eta^2}{4D}\right) =1 - \frac{2}{3} \frac{\eta^2}{4D},     
\end{equation}   
and the function $U$ is also a polynomial with the first order. 
We can conclude, that the sum of $c_1 M + c_2 U$ is also a polynomial with the first order. 
The general solution reads in this case  
\begin{equation} 
C(x,t) = \frac{1}{t^2} f(\eta) =  \frac{1}{t^2} \eta  e^{-\frac{\eta^2}{4D}} \cdot  
\left( \kappa_0 + \kappa_1  \frac{\eta^2}{4D} \right), 
\label{sol2}
\end{equation}   
where $ \kappa_0  $ and $ \kappa_1  $ are real constants.

Following the above argumentation, for $\alpha=n$, (where $n>2$) 
yields the solution of  
 \begin{equation} 
C(x,t) = \frac{1}{t^n} f(\eta) =  \frac{1}{t^n} \eta  e^{-\frac{\eta^2}{4D}} \cdot  
\left( \kappa_0 +\kappa_1  \frac{\eta^2}{4D}  + ... + \kappa_{n-1} \cdot \left[\frac{\eta^2}{4D}\right]^{n-1}   \right). 
\end{equation}   

 At this point we try to determine the concrete values of the coefficients $\kappa_n$. 
For the very first case, if $\alpha=1$, there is a single multiplicative constant multiplying the function $f(\eta)$. 

For $\alpha=2$, the situation is a little bit more complex. By reinserting the function 
$f(\eta)$ of  formula (\ref{sol2}), into the equation (\ref{ode1}) we get 
\begin{equation} 
f(\eta) =  \eta \cdot  e^{-\frac{\eta^2}{4D}} \cdot  
\kappa_0  \left[ 1 -  \frac{1}{6D} \eta^2   \right].  
\end{equation}   
If we incorporate the diffusion coefficient by rescaling the time, or if it is taken to be one $D=1$, we have for $ C(x,t) $
\begin{equation} 
C(x,t) = \frac{1}{t^2} f(\eta) =  \frac{1}{t^2} \eta  e^{-\frac{\eta^2}{4D}} \cdot  
\kappa_0  \left[  1 -  \frac{1}{6} \eta^2   \right].   
\end{equation}   

\begin{figure}[htp]
   \scalebox{0.45}{   {\includegraphics{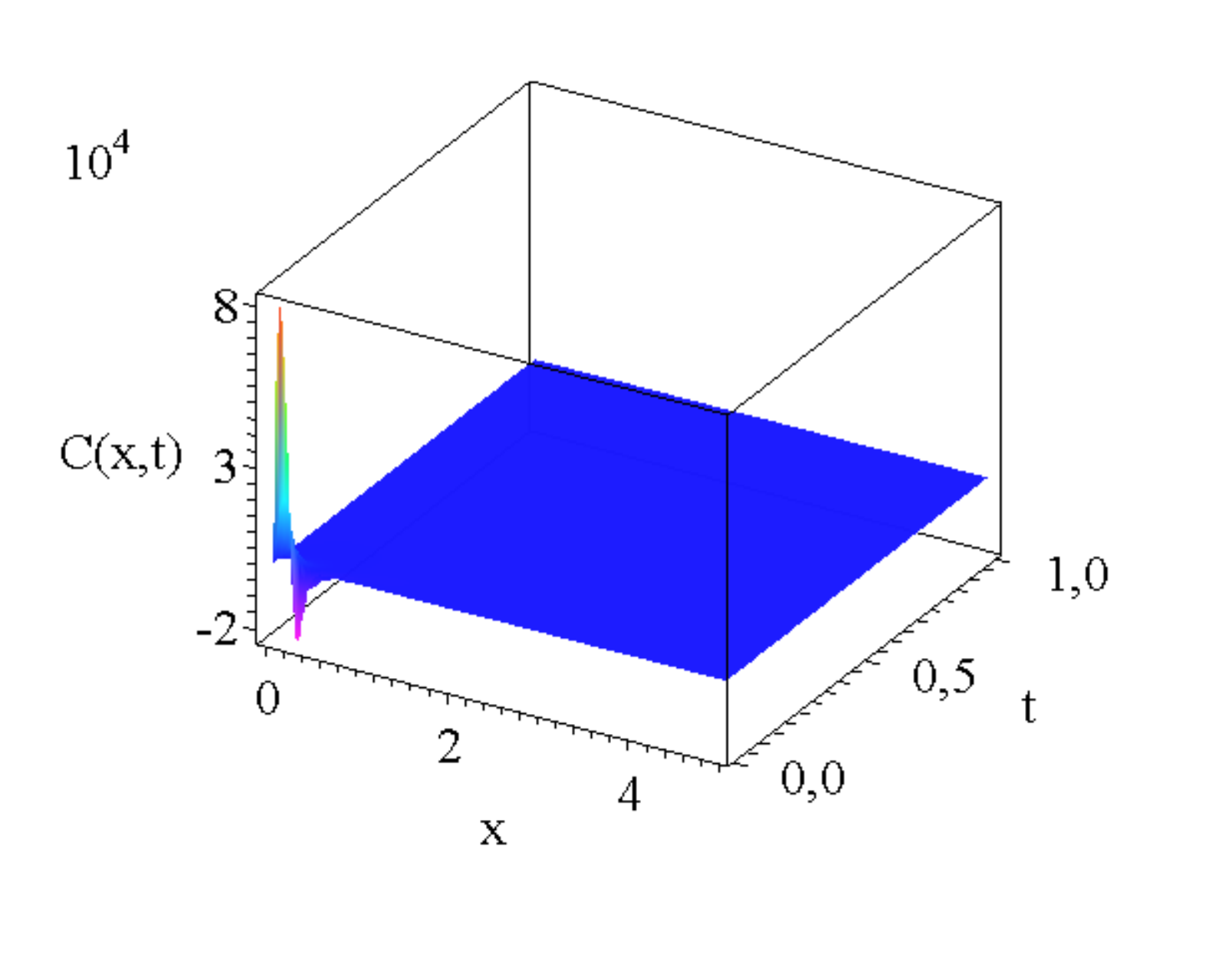}} } 
   \hspace*{0cm}    \scalebox{0.45}{   {\includegraphics{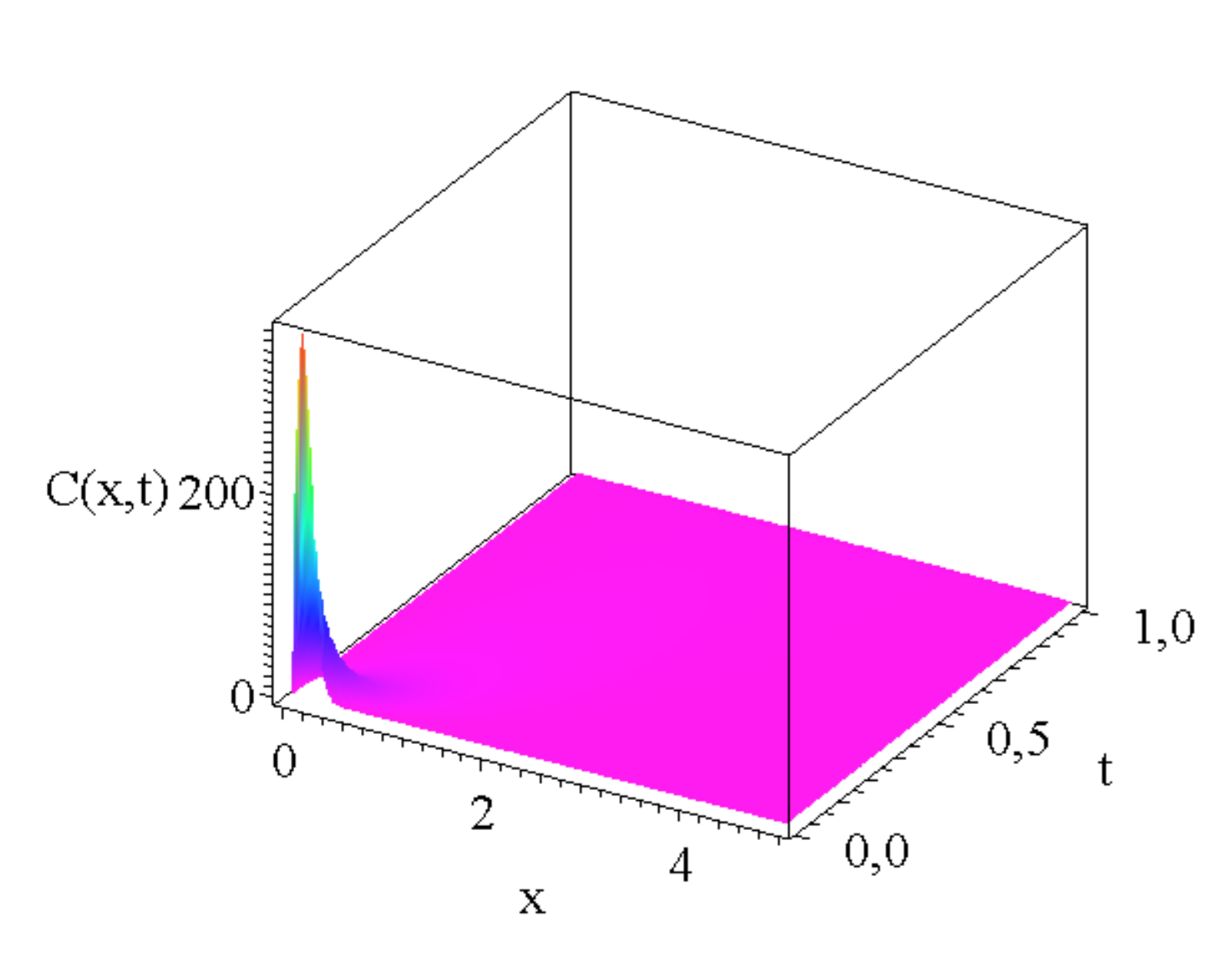}}}  \\
\hspace*{-2cm}     
  $\alpha = + 2 $     \hspace*{7cm} $\alpha = + 1 $  \\ 
   \scalebox{0.45}{   {\includegraphics{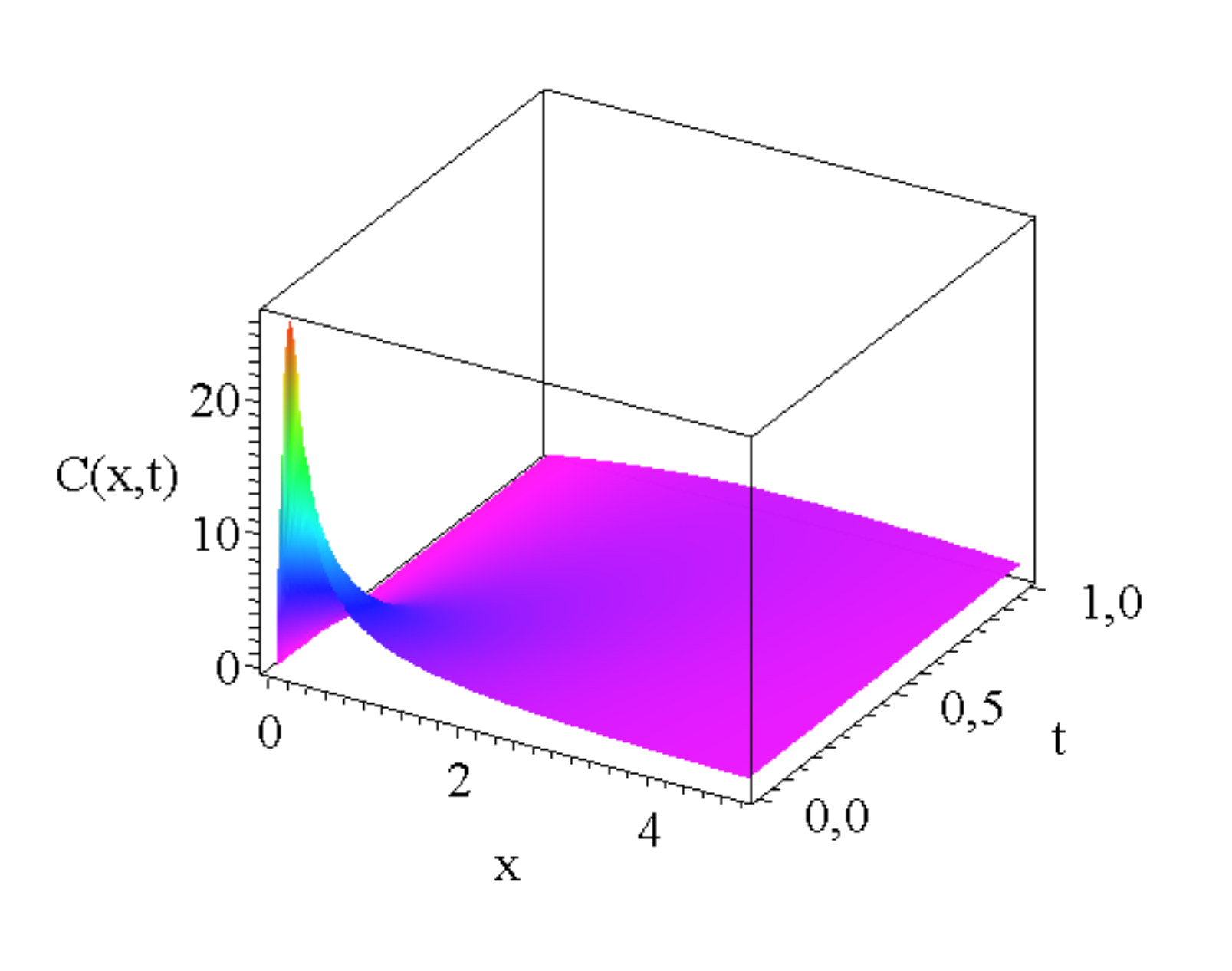}}} 
   \hspace*{0cm}    \scalebox{0.45}{   {\includegraphics{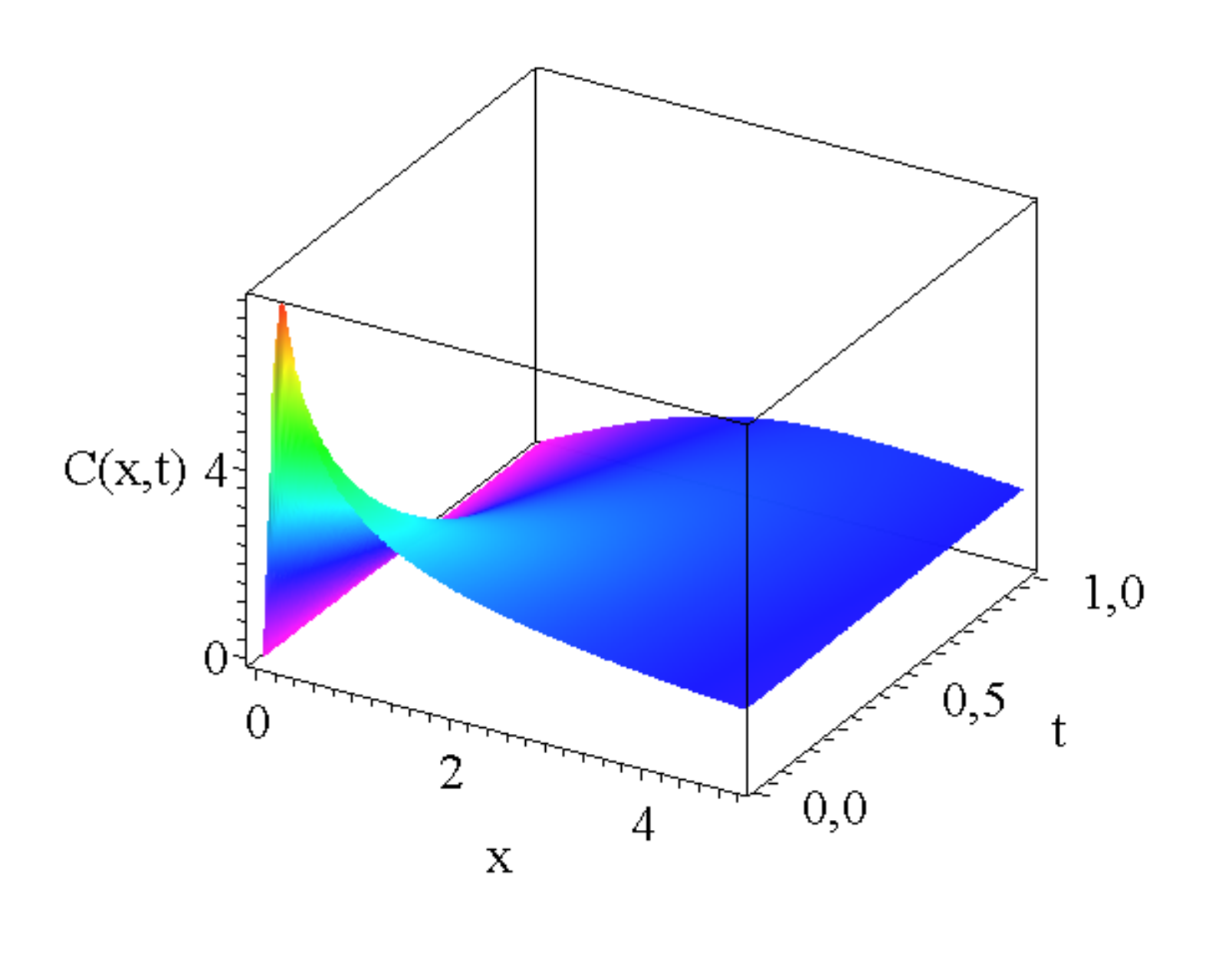}}} \\
    \hspace*{-3cm}  $\alpha = +\frac{1}{2} $     \hspace*{7cm} $\alpha = +\frac{1}{4}  $  \\
    \scalebox{0.45}{   {\includegraphics{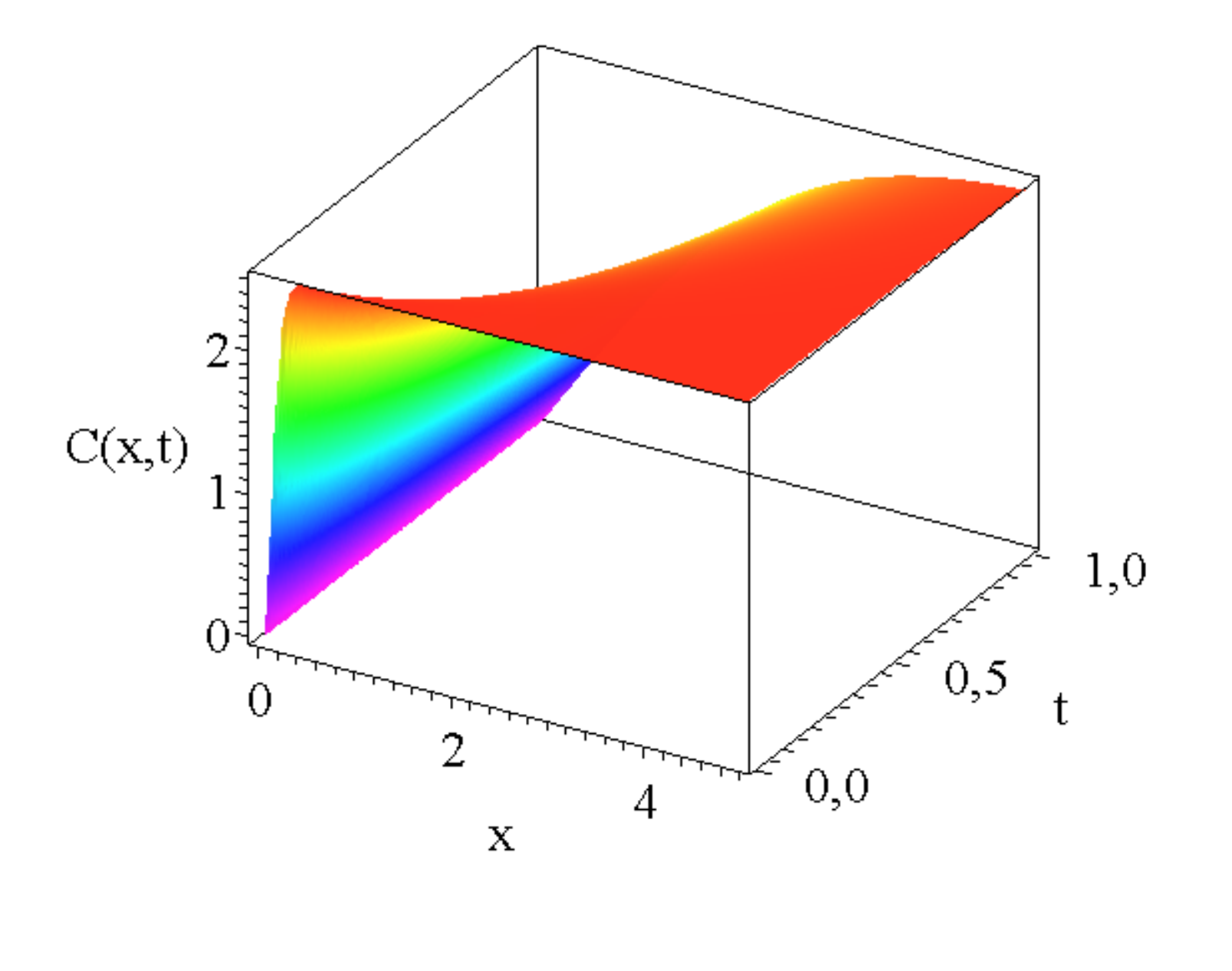}}} 
   \hspace*{0cm}    \scalebox{0.45}{   {\includegraphics{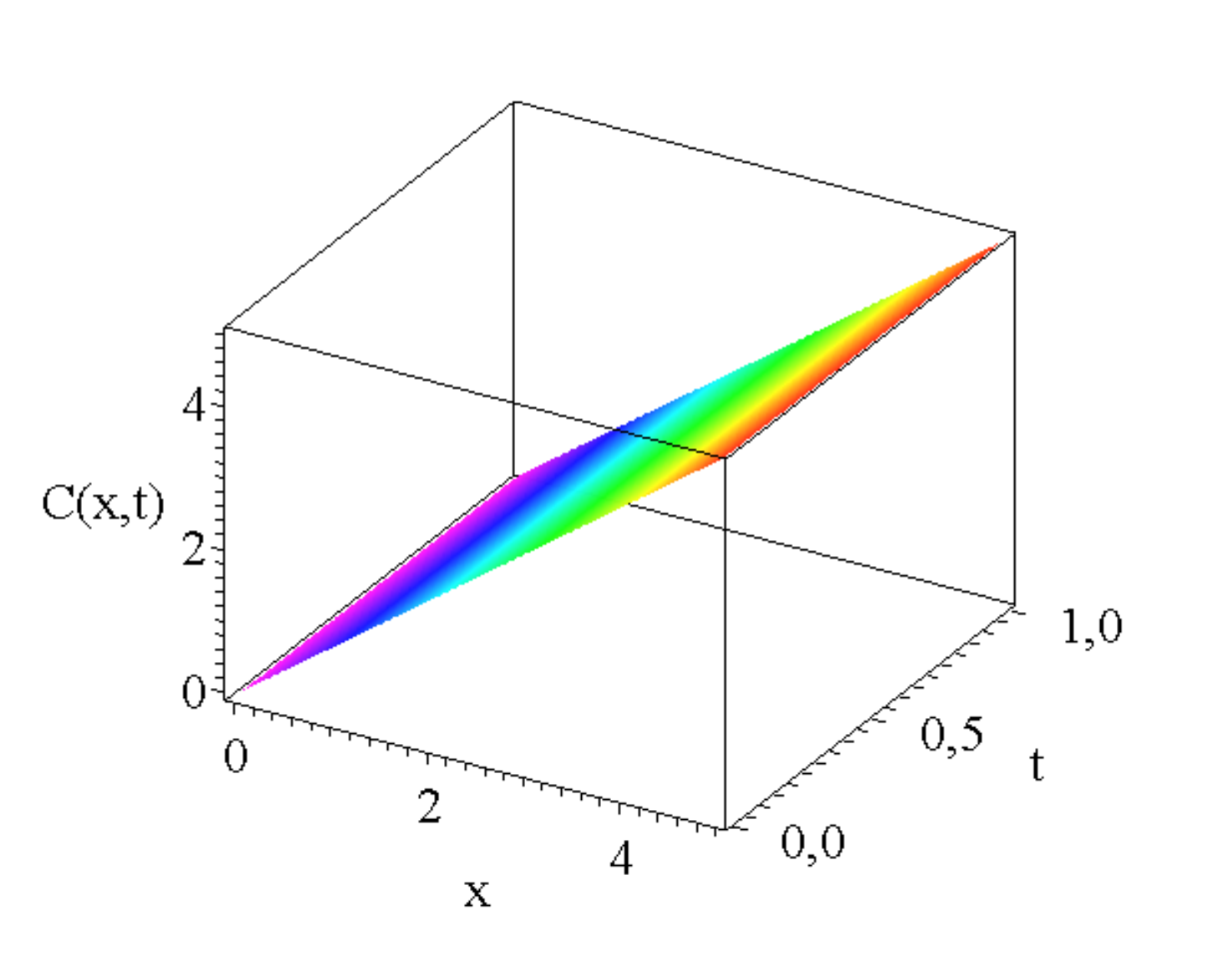}}} \\
    \hspace*{-2cm}  $\alpha = 0 $     \hspace*{7cm} $\alpha =- \frac{1}{2}   $  \\
\caption{ The total solutions $C(x,t)$ with the shape function of (\ref{f_eta2}) for six various $\alpha$ values.   
Additional parameters $D = 2, c_2 = 1, c_2 = 0$ are the same 
in all cases. Note, that for a better comparison the same ranges are taken 
for the spatial and temporal variables in all six graphs. }  
\label{fig_eta_time}
\end{figure}

 Figure 4 presents the final solutions of $C(x,t)$ evaluated from the shape function of 
Eq. (\ref{f_eta2}) for six different $\alpha$s. Note, that all positive $\alpha$s mean 
decaying solutions. The $\alpha = 0$ is the limiting case means an asymptotically  
constant solution, and negative $\alpha$s solutions diverge at large times. 

For $\eta \rightarrow 0$, the expression $e^{-\eta^2/(4D)}$ tends to one. By this, the function $f$ for given $x$, and large times decays like 
\begin{equation} 
 f \sim \frac{1}{t}  . 
\end{equation}
As a consequence for finite $x$, and given value of $\alpha > 0$ mentioned above, the expression $C(x,t)$ decays for sufficiently large times in the following way 
\begin{equation} 
 C(x,t) \sim \frac{1}{t^{(\alpha + 1)}} .  
\end{equation}

At last just for the sake of completeness, we mention that with the $t  = t_0 -t$ substitution we can get the so called blow-up solutions. The functional form of $(\ref{f_eta2})$ remains unchanged, and the graphs of the shape functions are changeless. Just the final $C(x,t)$ solutions go to infinity after a finite time for positive $\alpha$s. Solutions with  zero and negative  $\alpha$  values however remain unchanged.   Two of such solutions are visualized in the last, $5^{th}$ figure.  
\begin{figure}[htp]
     \scalebox{0.6}{   {\includegraphics{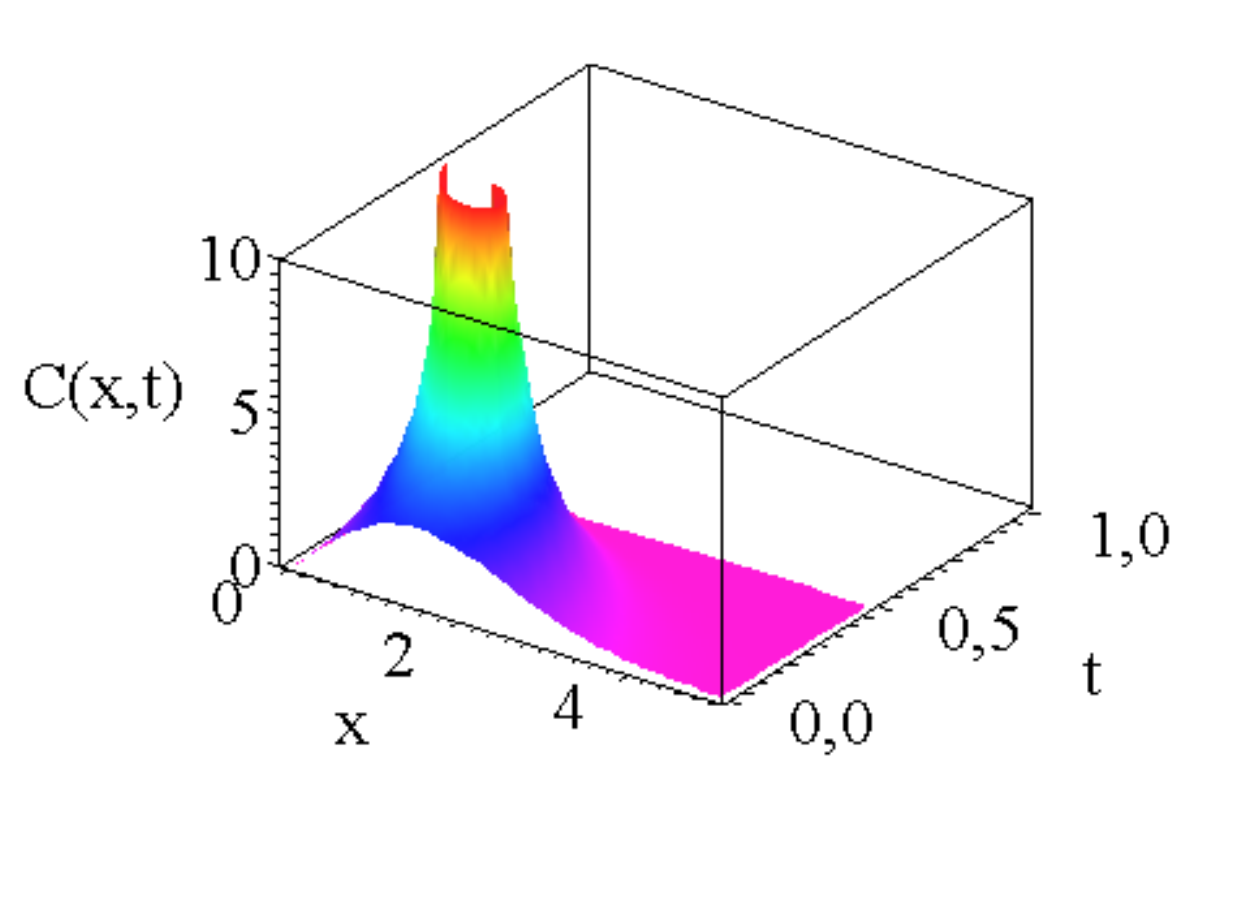}} } 
   \hspace*{0cm}    \scalebox{0.6}{   {\includegraphics{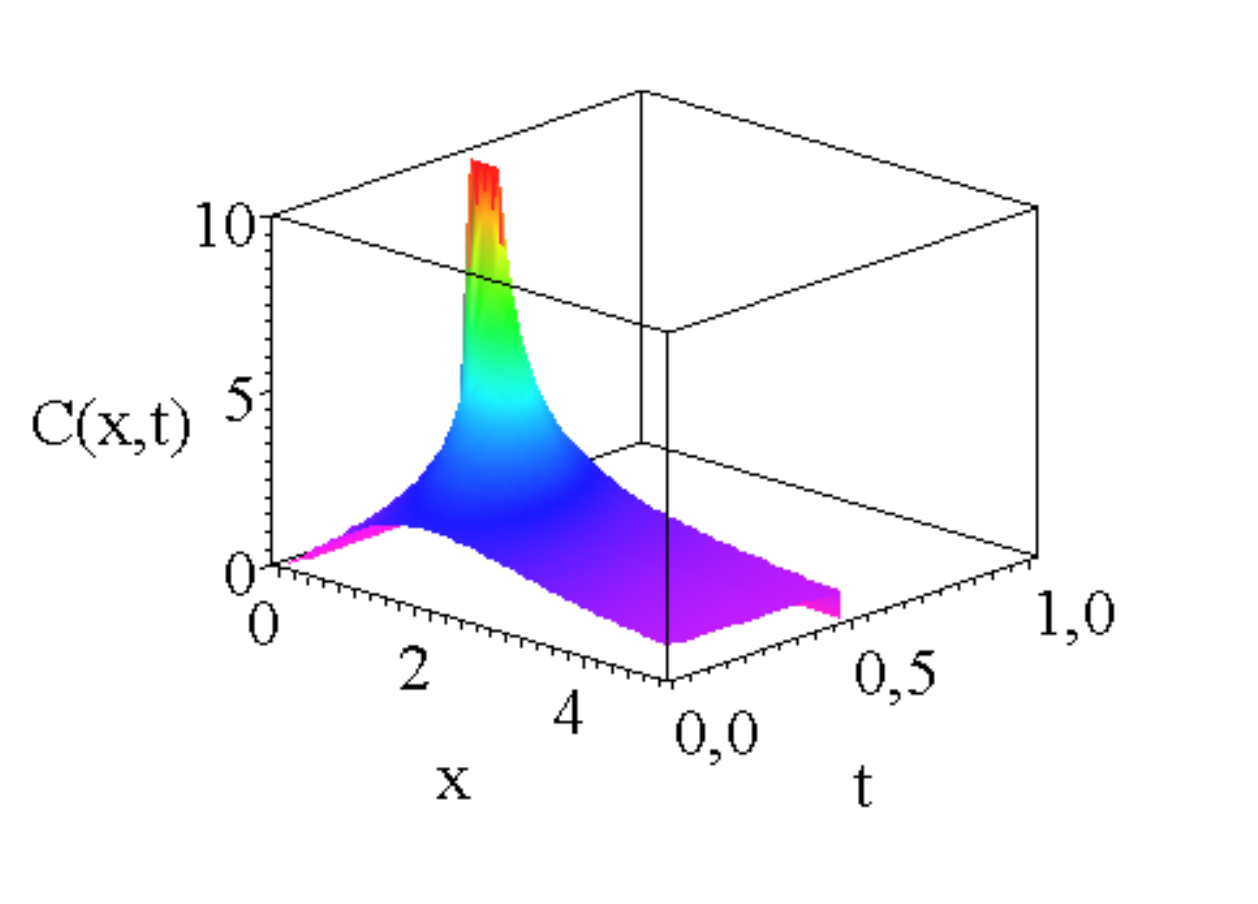}}}  \\    
   $\alpha = + 1 $     \hspace*{+5cm} $\alpha = + \frac{1}{2} $  \\ 
\caption{Two solutions from the {\it{blow-up}} kind, with the same parameters given above. } 
\end{figure}
We think that our exhaustive analysis in general helps the reader to understand the complex beauty of the solutions of PDEs especially the diffusion equation. 
The second aim of our study is, that these concrete results could attract the interest of the community of anomalous diffusion \cite{anomdif} or anomalous transport \cite{klages,KoChKl2005,KoBa2011,LiKrDe2018,GiKlSo2019}.  
\section{Summary} 
After a short historical summary of diffusion we presented analytic results obtained  
from three different Ans\"atze. First from the non-classical method of group invariant method, then from the 
traveling profile and finally the classical self-similar Ansatz.  
The results evaluated from the last trial function were analyzed in details, numerous 
formulas are given for different $\alpha \ge 0$ self-similar exponents which all mean 
physically relevant decaying solutions with different temporal asymptotics.  
Such results might exist deeply hidden  in intrinsic dynamics of certain systems. 
As limiting solution the $\alpha = 0$ was discussed in connection with fluid evaporation. 
Future work is in progress to perform comparable analysis among the three mentioned  
trial functions for non-linear diffusion equations as well. 
 A straightforward organic generalization is when both  $\alpha,\beta $ 
exponents can take arbitrary real numbers, it will be shown in future studies that such cases may be related to 
diffusion equations which have time-dependent diffusion coefficients.  
Investigation of processes where the diffusion coefficients have spatial dependence is also a future challenge. 
This kind of in-depth similarity analysis would be desirable and instructive for second oder wave equations, too. 
\section{Acknowledgment}
One of us (I.F. Barna) was supported by
the NKFIH, the Hungarian National Research
Development and Innovation Office. 

\end{document}